\documentstyle[multicol,aps,epsfig]{revtex}  % DON'T CHANGE

\begin{document}                % INITIALIZE - DONT CHANGE
\title{In-medium modifications of the $\pi\pi$ interaction in photon-induced reactions}
\author{J.G~Messchendorp,$^1$ S.~Janssen,$^1$ M.~Kotulla,$^1$ 
	J.~Ahrens,$^2$ J.R.H.~Annand,$^2$ R.~Beck,$^2$ 
	F.~Bloch,$^3$ G.~Caselotti,$^2$ L.~Fog,$^4$ D.~Hornidge,$^2$ 
	B.~Krusche,$^3$ W.~Langg\"artner,$^1$ J.C.~McGeorge,$^4$ 
	I.J.D.~MacGregor,$^4$ K.~Mengel,$^1$ V.~Metag,$^1$ R.~Novotny,$^1$ 
        R.O.~Owens,$^4$ M.~Pfeiffer,$^1$ S.~Sack,$^1$ R.~Sanderson,$^4$ 
        S.~Schadmand~$^1$}
\address{$^1$II. Physikalisches Institut,
         University of Gie{\ss}en, 
         D-35392 Gie{\ss}en, Germany}
\address{$^2$Institut f\"ur Kernphysik,
         Johannes-Gutenberg-Universit\"at Mainz,
         D-55099 Mainz, Germany}
\address{$^3$Department of Physics and Astronomy,
	 University of Basel, CH-4056 Basel,
         Switzerland}
\address{$^4$Department of Physics and Astronomy, University of Glasgow, 
         Glasgow, G128QQ, United Kingdom}
\maketitle
\begin{abstract}
Differential cross sections of the reactions $(\gamma,\pi^\circ\pi^\circ)$ 
and $(\gamma,\pi^\circ\pi^{+/-})$ have been measured for several
nuclei ($^1$H,$^{12}$C, and $^{\rm nat}$Pb) at 
an incident-photon energy of $E_{\gamma}$=400-460~MeV at the tagged-photon facility at MAMI-B 
using the TAPS spectrometer. A significant
nuclear-mass dependence of the $\pi\pi$ invariant-mass distribution is found in the 
$\pi^\circ\pi^\circ$ channel. This dependence is not observed in the $\pi^\circ\pi^{+/-}$ 
channel and is consistent with an in-medium modification of the 
$\pi\pi$ interaction in the $I$=$J$=0 channel. The data are compared to $\pi$-induced 
measurements and to calculations within a chiral-unitary approach.
\end{abstract}
% \pacs{13.60.L} \pacs{21.65.+f} \pacs{11.30.Rd}

\begin{multicols}{2}

% Introduction......

One of the challenges in nuclear physics is to study
the properties of hadrons and the modification of these 
properties when the hadron is embedded in a nuclear 
many-body system. Although much has been learned about the properties
of hadrons in free space, there is a lack of information
for particles in a dense environment.
In this Letter, an experiment is described which has measured
correlated pion pairs photoproduced on nuclei in the scalar-isoscalar 
$J$=$I$=0 channel, also known as the $\sigma$ mode.
In Ref.~\cite{gro00} the $\sigma$ meson is
identified as the $f_0$(400-1200). The large natural 
width in free space of $\Gamma$=400-500~MeV~\cite{oll98} makes it doubtful 
that this particle is a mesonic state, and has initiated many discussions on its 
nature. An in-medium study of the $I$=$J$=0 
channel could provide a better insight into the nature of the 
$\sigma$ meson.

Within some theoretical approaches 
of quantum chromodynamics (QCD)~\cite{hat99,sch00,cha01}, 
the $\sigma$ is treated as a pure $q\bar q$ state ($J^P$=0$^+$) and 
regarded as the chiral partner of the pion ($J^P$=0$^-$).
Chiral symmetry is spontaneously broken in the QCD vacuum, resulting in a 
mass difference between the pion and the $\sigma$. 
For large baryon densities, it is predicted that chiral symmetry is partially 
restored, leading to a degeneracy in mass of the pion and the $\sigma$.
Since the pion approximates a Goldstone boson, the pion mass is not expected 
to change dramatically with increasing nuclear density $\rho$. 
Hence, these models predict a significant drop in the mass of the $\sigma$.
A measurement of the in-medium $\sigma$$\rightarrow\pi\pi$ mass distribution 
might be essential for the understanding of the mechanism of chiral-symmetry 
breaking.

Alternatively, the in-medium $\sigma$ mode can be considered to be a resonant state
of two pions~\cite{sch88,chi98,ose01,roc02}. In vacuum, the $\pi\pi$ system is mildly 
attractive. However, in the nuclear medium the $\pi\pi$ interaction strength 
could increase, thereby changing width and pole position of the resonant state.
Experimental data on the density dependence of pion-pair interactions in
the nuclear medium can provide evidence for this phenomenon.

% pion-induced experiments

The first measurement of the in-medium $\pi\pi$ mass was obtained by a
pion-induced experiment by the CHAOS collaboration~\cite{bon96,bon99,bon00}. 
A rising accumulation of strength at low $\pi^+\pi^-$ mass was observed with
increasing nuclear mass whereas such 
an enhancement was not seen in the $\pi^+\pi^+$-mass distributions. This effect was 
interpreted as a signature for an in-medium modification of the $\pi\pi$ interaction 
in the $I$=$J$=0 channel. A similar effect was found by a pion-induced experiment 
of the Crystal Ball collaboration~\cite{sta00} where a nuclear-mass dependence of 
the $\pi^\circ\pi^\circ$-mass distribution was observed. 

% intro photon-induced experiments

For the interpretation of the previously described pion-induced measurements
two issues have to be addressed. The first one results from the final-state 
interactions, rescattering and absorption, of the pions. Such effects distort
the actual $\pi\pi$-mass measurement. To minimize pion final-state interactions,
the incident-beam energy was chosen such that the energies of the outgoing pions 
were small, thereby maximizing their mean-free path.    
The second issue is the strong interaction of the initial-state
pion with the medium. As a result, only the
surface of the nucleus is probed, leading to a small effective nuclear
density. The authors of Ref.~\cite{vac99} estimate an average density of 24\%
of the interior nuclear density $\rho_0$=0.17~fm$^{-3}$ for $^{40}$Ca. 
It was therefore proposed to produce in-medium $\pi\pi$ pairs
with electromagnetic probes, which illuminate the complete nucleus, and
lead to a larger effective density. 

% This experiment....

In this Letter, we present measurements of
$A(\gamma,\pi^\circ\pi^\circ)$ and $A(\gamma,\pi^\circ\pi^{+/-})$ for $A$=$^1$H, $^{12}$C,
and $^{\rm nat}$Pb. These measurements allow to study the different
$\pi\pi$-isospin states at average effective densities of 35\% ($^{12}$C) to
65\% ($^{208}$Pb)~\cite{roc02b} of $\rho_0$ and are statistically superior
to previously published data on photon-induced double-pion 
production~\cite{wol00,lan01}.
Data are presented for an incident-photon energy of
$E_{\gamma}$=400-460~MeV. 
The centroid of this interval corresponds to the same center-of-mass 
energy as was used in the pion-induced experiments, enabling a direct 
comparison and minimizing the effect of final-state interactions 
of the two pions with the medium. 

% General experimental issues

The experiment was performed at the photon-beam facility at MAMI-B.
Tagged photons~\cite{ant91,hal96} were produced with energies between 
200 and 820~MeV. The beam intensity in the energy range of interest, 
$E_{\gamma}$=400-460~MeV, was 10$^7$~s$^{-1}$ with a photon-energy 
resolution of about 2~MeV. After collimation, the
photon beam was transported to a nuclear target in an evacuated beam line.
A series of measurements were carried out using
liquid-hydrogen, carbon, and lead targets with thicknesses of 10~cm, 
2.5~cm, and 5~mm, respectively. The photon-conversion ($\gamma\rightarrow e^+e^-$) 
probability for all targets is smaller than 10\%.

\begin{figure}
\vspace*{-2.2cm}
\hspace*{-1.8cm}
\psfig{figure=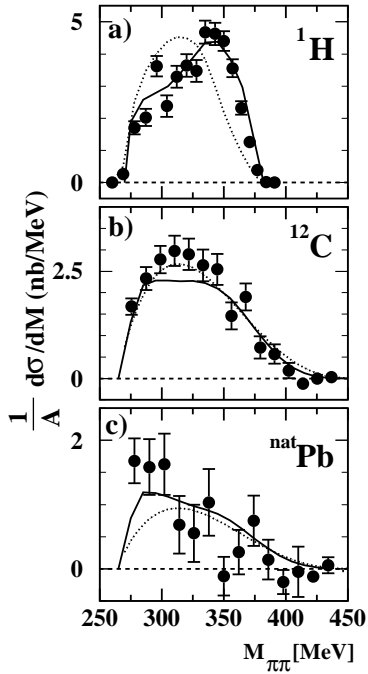,width=11.0cm}
\vspace*{-0.5cm}
\caption{Differential cross sections of the reaction $A(\gamma,\pi^\circ\pi^\circ)$ with
$A$=$^1$H,$^{12}$C,$^{\rm nat}$Pb for incident photons in the energy range
of 400-460~MeV (solid circles).
Error bars denote statistical uncertainties and the curves are explained in the text.}
\label{mpi0pi0-spectrum}
\end{figure}

The angles and energies of the pions 
were measured using the TAPS photon spectrometer~\cite{gab94}.
In this experiment, the TAPS detector consisted of 510 hexagonal
BaF$_2$ scintillators. Sixty-two crystals, arranged in an 8$\times$8 matrix,
formed a TAPS block. Six blocks were mounted coplanar with
the target at a distance of 55~cm and polar angles of $\pm$55$^{\circ}$,
$\pm$105$^{\circ}$ and $\pm$155$^{\circ}$ with respect to the photon-beam direction.
The remaining 138 BaF$_2$ crystals were arranged in a rectangular forward
wall which covered polar angles between 5$^{\circ}$ and 38$^{\circ}$. 
The complete setup covered $\approx$40\% of the total 
solid angle. Photons and charged pions were identified by exploiting the 
time-of-flight information of each detector. 
A 5~mm thick plastic scintillator was placed in front of each crystal
to differentiate between neutral and charged particles.

% Analyses steps

Neutral pions were identified by an invariant-mass analysis of the two decay photons.
The two-photon invariant-mass resolution ($\sigma$) for $\pi^\circ$ is 5.7\%. 
A kinematic fit was applied to improve the pion-energy resolution~\cite{kor00}.
For the identification of the $A(\gamma,\pi^\circ\pi^\circ)$ reaction, all four 
final-state photons were registered in the detector. The two-$\pi^\circ$ invariant-mass 
($M_{\pi^\circ\pi^\circ}$) resolution ($\sigma$) varies between 2.0\% and 2.5\% in the 
incident-photon energy range of interest.

The capability to detect and distinguish neutral from
charged pions is essential for comparing pion pairs of different isospin.
Charged pions from $A(\gamma,\pi^\circ\pi^{+/-})$ 
were selected by exploiting the information on the time-of-flight of the charged pion 
relative to the one of the photons of the $\pi^\circ$ decay and its deposited energy
in the BaF$_2$ crystals~\cite{jan02}. 
%In the analysis of the data discussed in this
%Letter, the forward wall was not used for the detection of charged pions.  
\noindent Since the TAPS detector does not include a 
magnetic field, positively charged particles cannot be discriminated from 
negatively charged particles. 
The two-pion mass resolution ($\sigma$) in the $\pi^\circ\pi^{+/-}$ 
channel is $<$3.3\%. 

The dominant reaction mechanism in $A(\gamma,\pi^\circ\pi^\circ)$ and 
$A(\gamma,\pi^\circ\pi^{+/-})$ channels is the quasi-free production
on the constituent nucleons. Under this assumption, the undetected recoil nucleon was 
deduced from the incident photon energy and the momenta of the final-state pions. 
Its reconstructed-mass distribution was found to be consistent with Monte-Carlo 
simulations using a quasi-free event generator.
The background of the $\eta\rightarrow 3\pi^\circ$ production channel does not contribute, 
since the incident-photon energy of $E_{\gamma}$=400-460 is below the $\eta$-production 
threshold. 

Cross sections were deduced from the yield of the $\pi\pi$ events divided by the
thickness of the targets, the photon flux, efficiencies, geometrical acceptances,
and the branching ratio $\pi^\circ\rightarrow \gamma\gamma$. The intensity of the photon
beam was determined by counting the post-bremsstrahlung electrons in the 
focal plane of the tagger. 
The loss of photon intensity due to collimation was measured with a 100\%-efficient 
BGO detector which was moved into the photon beam at lowered beam intensity.
The geometrical acceptance and inefficiencies due to cuts and thresholds were 
deduced from a Monte-Carlo simulation based on GEANT3\cite{geant3} libraries and 
an event generator assuming a quasi-free production mechanism. The generator was 
modified such that energy and angular distributions of the final-state particles
agreed well with the observed distributions~\cite{jan02}. The obtained acceptance 
was found to be typically 0.2-0.4\%.

% The actual results...

The measured $M_{\pi^\circ\pi^\circ}$-mass distributions for incident-photon energies of
$E_{\gamma}$=400-460~MeV are shown in Fig.~\ref{mpi0pi0-spectrum}.
A strong increase in strength towards small $M_{\pi^\circ\pi^\circ}$ with increasing $A$
is observed.
The dotted curves in Fig.~\ref{mpi0pi0-spectrum}
indicate phase-space distributions determined by the Monte-Carlo model.
The experimentally observed peak position for
$A$=$^1$H (a) lies higher than the phase-space prediction whereas for
$A$=$^{12}$C (b) the measured mass
distribution is compatible with phase space. For $A$=$^{\rm nat}$Pb (c), the data disagree with
phase space with a probability of more than 99.8\%. Most of the observed strength lies 
below the peak of the phase-space distribution. A similar, but less pronounced, effect 
has been observed in pion-induced reactions $A(\pi^-,\pi^\circ\pi^\circ)$~\cite{sta00} at a 
comparable center-of-mass energy.  The experimentally determined angular distributions in the 
$A(\gamma,\pi^\circ\pi^\circ)$ reaction of the $\pi^\circ\pi^\circ$ center-of-mass 
system are found to be isotropic~\cite{jan02}
and are compatible with $J$=0, supporting the conclusion that a significant 
$A$ dependence is found in the $\pi\pi$ $I$=$J$=0 channel in photon-induced reactions.

The solid curves in Fig.~\ref{mpi0pi0-spectrum} are predictions by Roca et al.~\cite{roc02}.
Here, the meson-meson interaction in the scalar-isoscalar channel is studied in the framework
of a chiral-unitary approach at finite baryonic density. The model dynamically generates the 
$\sigma$ resonance, reproducing the meson-meson phase shifts in vacuum and accounts 
for the absorption of the pions in the nucleus. 
The data are described well by the model considering a theoretical uncertainty 
of 20\%~\cite{roc02b}. 
It qualitatively predicts a mass shift as observed in the data. The basic ingredient 
driving this shift is the p-wave interaction of the pion with the baryons in the medium, 
resulting in an in-medium modification of the $\pi\pi$ interaction. 
A similar calculation~\cite{vac99} is not able to describe the observed $A$-dependence effect 
in the $A(\pi^-,\pi^\circ\pi^\circ)$ data~\cite{sta00}, which might be due to the 
interaction of the initial-state pion.

In order to compare the TAPS results with the pion-induced measurements by the CHAOS collaboration
($A(\pi^+,\pi^+\pi^-)$), the composite ratio $C_{\pi\pi}$ is introduced~\cite{bon00} 
\begin{eqnarray}
C_{\pi\pi}(Pb/C)=\frac{(d\sigma (Pb)/dM)/\sigma (Pb)}{(d\sigma (C) /dM)/\sigma (C)}.\nonumber
\end{eqnarray}

\begin{figure}
\vspace*{-1.0cm}
\hspace*{0.0cm}
\psfig{figure=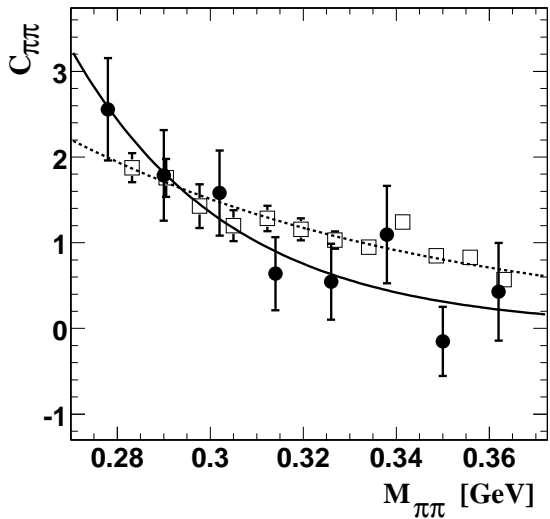,width=8.cm}
\caption{The composite ratio $C_{\pi\pi}$ for A($\gamma,\pi^\circ\pi^\circ)$ 
(full symbols) compared to $A(\pi^+,\pi^+\pi^-)$ (open squares) obtained by the CHAOS
collaboration~\protect\cite{bon96,bon99,bon00}. The curves are second-order polynomial fits 
through the data.}
\label{mpi0pi0-ratio}
\end{figure}

\noindent The results are shown in Fig~\ref{mpi0pi0-ratio}.
The photon-induced $A(\gamma,\pi^\circ\pi^\circ)$ data (solid circles) are 
compared to the pion-induced $A(\pi^+,\pi^+\pi^-)$ measurement by the 
CHAOS collaboration~\cite{bon96,bon99,bon00} 
(open squares). The solid and dashed curves represent empirical 
second-order polynomial fits through the photon-induced and pion-induced data, respectively. 
In both cases, an increase in strength towards small $M_{\pi\pi}$ masses is observed. 
This increase is stronger in $A(\gamma,\pi^\circ\pi^\circ)$ than 
in $A(\pi^+,\pi^+\pi^-)$ reactions, 
which could be related to photons probing the entire nucleus leading to 
larger effective densities than with pion beams. 

\begin{figure}
\vspace*{-2.2cm}
\hspace*{-1.8cm}
\psfig{figure=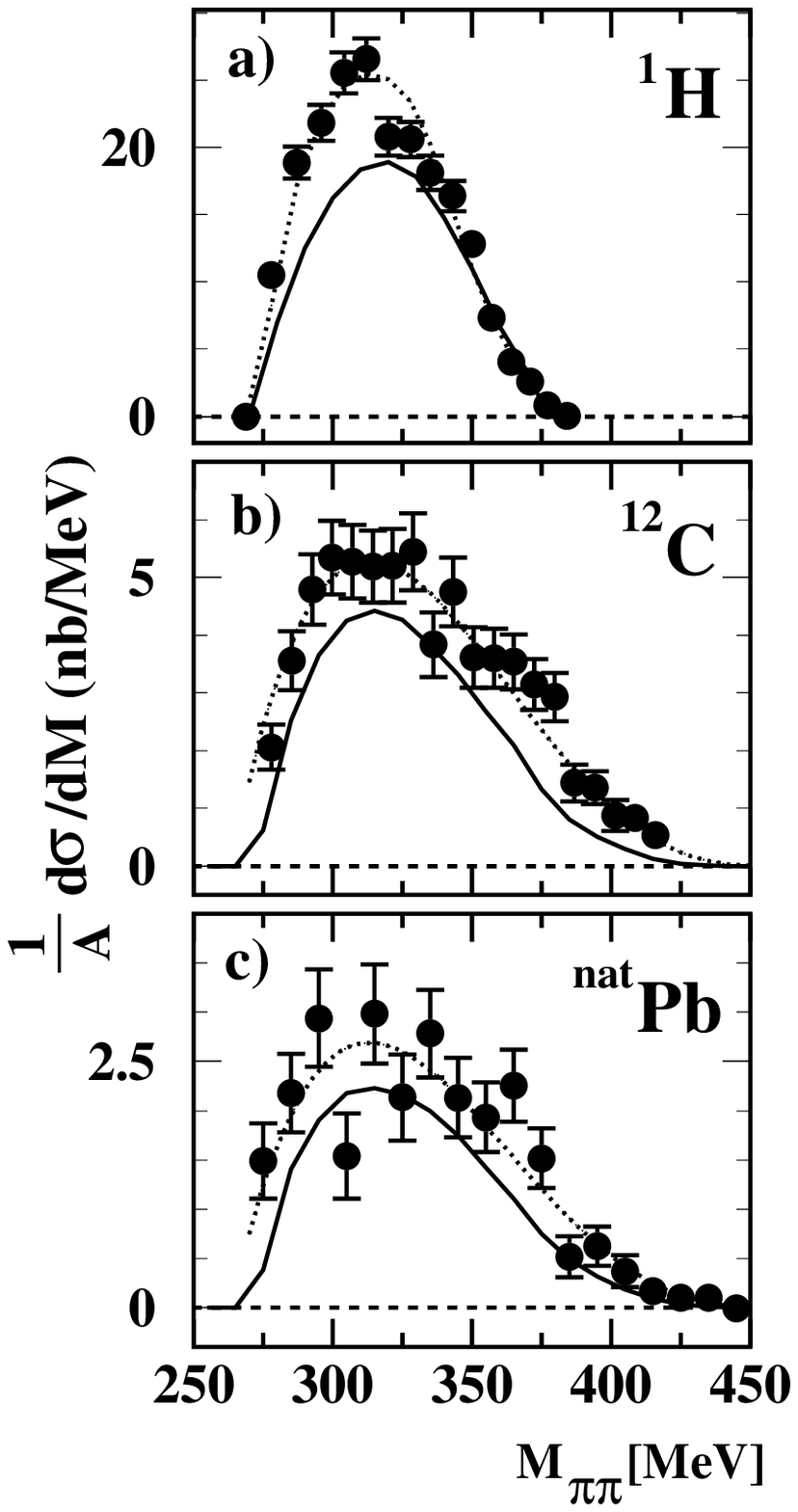,width=11.0cm}
\vspace*{-0.5cm}
\caption{Differential cross section of the reactions $p(\gamma,\pi^\circ\pi^+)$ (a) and
$A(\gamma,\pi^\circ\pi^{+/-})$ with
$A$=$^{12}$C,$^{\rm nat}$Pb (b,c) for incident photons in the energy range of 
400-460~MeV (solid circles). 
Error bars denote statistical uncertainties and the curves are explained in the text.}
\label{mpi0pip-spectrum}
\end{figure}

% Different isospin channels

To study the nuclear-mass dependence of the double-pion mass in a different isospin channel
than $I$=0, we have concurrently measured differential cross sections of the reactions 
$A(\gamma,\pi^\circ\pi^{+/-})$. The same energy interval of $E_{\gamma}$=400-460~MeV was chosen. 
The results for $A$=$^1$H,$^{12}$C and $^{\rm nat}$Pb are depicted in Fig.~\ref{mpi0pip-spectrum}. 
The data do not show an $A$ dependence in shape as was observed in the corresponding 
$M_{\pi^\circ\pi^\circ}$ distributions. For all targets, the data follow the phase-space 
distributions depicted as dotted curves, indicating that significant in-medium
effects in the isospin $I$=1 channel are not observed. The solid curves represent 
predictions by Roca et al.~\cite{roc02b} and are performed in a similar framework
as the model for $M_{\pi^\circ\pi^\circ}$ distributions~\cite{roc02}.
The model underestimates the experimentally determined cross sections by $\approx$20\% for all
nuclei, while describing the shape of the data rather accurately. Since the $\sigma$ resonance 
does not couple to $\pi^\circ\pi^{+/-}$, the model shows no
shift in strength towards smaller $M_{\pi\pi}$ masses with increasing $A$.

\begin{figure}
\vspace*{-3.0cm}
\hspace*{-2.0cm}
\psfig{figure=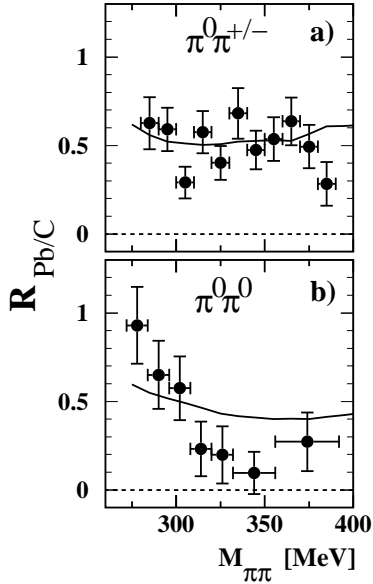,width=11cm}
\vspace*{-0.5cm}
\caption{Ratios between the differential cross sections for $A$=$^{\rm nat}$Pb and $A$=$^{12}$C
for $A(\gamma,\pi^\circ\pi^{+/-})$ (a) and $A(\gamma,\pi^\circ\pi^\circ)$ (b).
The solid curves represent predictions by Roca et al.\protect\cite{roc02,roc02b}.}
\label{mpipi-ratio}
\end{figure}

Figure~\ref{mpipi-ratio} shows the ratio $R_{\rm Pb/C}$ between the 
differential cross sections per nucleon for $A$=$^{\rm nat}$Pb and $A$=$^{12}$C of the
reactions $A(\gamma,\pi^\circ\pi^{+/-})$ (a) and $A(\gamma,\pi^\circ\pi^\circ)$ (b)
up to $M_{\pi\pi}$ masses of 400~MeV.
The experimentally determined ratio $R_{\rm Pb/C}$ for the $\pi^\circ\pi^{+/-}$ reaction is 
found to be flat, indicating that final-state interactions, absorption, and rescattering
of the individual pions with the medium do not modify the shape in the mass distribution
significantly. The model of Roca et al.~\cite{roc02b} supports this conclusion as can be
observed from the solid curve.
A significant in-medium shape effect is observed in the ratio $R_{\rm Pb/C}$ for 
the $\pi^\circ\pi^\circ$ channel as depicted in Fig.~\ref{mpipi-ratio} (b). 
Since an in-medium modification is not seen in the $\pi^\circ\pi^{+/-}$ reaction, 
this effect cannot be explained by $A$ dependencies in the production mechanism and
final-state interactions of the individual pions with the medium. 
%The model of Roca et al.~\cite{roc02} qualitatively explains the in-medium
%effect by a modification of the $I$=$J$=0 $\pi\pi$ interaction. 
The prediction by Roca et al.~\cite{roc02}
with a theoretical uncertainty of 10\%~\cite{roc02b} is depicted 
as the solid curve in Fig.~\ref{mpipi-ratio} (b).
%however not
%sufficiently to describe the data quantitatively as shown by a comparison with the 
%solid curve in Fig.~\ref{mpipi-ratio} (b).

In conclusion, we have observed an effect consistent with a significant in-medium 
modification in the $A(\gamma,\pi^\circ\pi^\circ)$ ($I$=$J$=0) channel.
For the first time,
the $A$ dependence of the $\pi\pi$-mass distributions in photon-induced
reactions on nuclei has been measured. 
With increasing $A$, the strength in these distributions is shifting towards
smaller invariant masses. Earlier measurements using pion beams found a similar, but less
pronounced effect. Photon-induced experiments have the advantage that initial-state interactions
are absent and larger effective densities can be reached which enhance 
in-medium effects. The distortion of the $\pi\pi$-mass distribution due to $A$ 
dependencies in the production mechanism and final-state 
interactions of the individual pions with the constituents of the nucleus have been 
studied by measuring the $\pi^\circ\pi^{+/-}$ mass distribution concurrently. 
A significant in-medium effect was not observed. 
According to Roca et al.~\cite{roc02}, the modification observed in the 
$\pi^\circ\pi^\circ$-mass distributions can be attributed to a change of the 
$\pi\pi$ interaction.
The comparison with the experimental data hints at the nature of the $\sigma$ meson
as a $\pi\pi$ resonance. It would be most desirable to confront this observation
with QCD models which treat the $\sigma$ as a $q\bar q$ state and explicitly take
chiral-symmetry restoration into account.

The authors gratefully acknowledge the outstanding support of the accelerator
group of the Mainz Microtron MAMI, as well as the other technicians and
scientists of the Institut f\"ur Kernphysik at the Universit\"at Mainz.
We thank E.~Oset, M.J.~Vicente Vacas, and L.~Roca for making their calculations 
available to us prior to publication. We also acknowledge many fruitful discussions
with W.~Cassing, R.~Meier, B.~Nefkens, and E.~Oset. This work was supported by Deutsche 
Forschungsgemeinschaft (SFB 201), the U.K. Engineering and Physical 
Sciences Research Council, and Schweizerischer Nationalfond.

\end{multicols}


\begin{references}
 \bibitem{gro00} D.E.~Groom et al., Eur. Phys. J. {\bf C15}, 1 (2000).
 \bibitem{oll98} J.A.~Oller et al., Phys. Rev. Lett. {\bf 80}, 3452 (1998).
% \bibitem{oll99} J.A.~Oller et al., Phys. Rev. {\bf D59}, 074001 (1999).
 \bibitem{hat99} T.~Hatsuda et al., Phys. Rev. Lett. {\bf 82}, 2840 (1999).
 \bibitem{sch00} P.~Schuck et al., Proceedings of the International Workshop XXVIII
                 on Gross Properties of Nuclei and Nuclear Excitations, Hirschegg, 73 (2000).
 \bibitem{cha01} G.~Chanfray, Nucl. Phys. {\bf A685}, 328c (2001).
% \bibitem{rap99} R.~Rapp et al., Phys. Rev. Lett. {\bf 82}, 1827 (1999).
 \bibitem{sch88} P.~Schuck, W.~N\"orenberg and G.~Chanfray, Z. Phys. {\bf A330}, 119 (1988).
 \bibitem{chi98} H.C.~Chiang et al., Nucl. Phys. {\bf A644}, 77 (1998).
 \bibitem{ose01} E.~Oset et al., nucl-th/0112033, 2001.
 \bibitem{roc02} L.~Roca et al., Phys. Lett. {\bf B541}, 77 (2002).
 \bibitem{bon96} F.~Bonutti et al., Phys. Rev. Lett. {\bf 77}, 603 (1996).
 \bibitem{bon99} F.~Bonutti et al., Phys. Rev. {\bf C60}, 018201 (1999).
 \bibitem{bon00} F.~Bonutti et al., Nucl. Phys. {\bf A677}, 213 (2000).
 \bibitem{sta00} A.~Starostin et al., Phys. Rev. Lett. {\bf 85}, 5539 (2000).
 \bibitem{vac99} M.J.~Vicente Vacas and E.~Oset, Phys. Rev. {\bf C60}, 064621 (1999).
 \bibitem{roc02b} L.~Roca, E.~Oset, and M.J.~Vicente Vacas, private communications.
% \bibitem{bra95} A.~Braghieri et al., Phys. Lett. {\bf B363}, 46 (1995).
% \bibitem{zab97} A.~Zabrodin et al., Phys. Rev. {\bf C55}, R1617 (1997).
% \bibitem{zab99} A.~Zabrodin et al., Phys. Rev. {\bf C60}, 055201 (1999).
% \bibitem{kle00} V.~Kleber et al., Eur. Phys. J. {\bf A9}, 1 (2000).
 \bibitem{wol00} M.~Wolf et al., Eur. Phys. J. {\bf A9}, 5 (2000).
 \bibitem{lan01} W.~Langg\"artner et al., Phys. Rev. Lett. {\bf 87}, 052001 (2001).
 \bibitem{ant91} I.~Anthony et al., Nucl. Inst. Meth. {\bf A301}, 230 (1991).
 \bibitem{hal96} S.J.~Hall et al., Nucl. Instr. and Meth. {\bf A368}, 698 (1996).
 \bibitem{gab94} A.R.~Gabler et al., Nucl. Instr. and Meth. {\bf A346},
168 (1994).
 \bibitem{kor00} K.~Korzecka, T.~Matulewicz, Nucl. Instr. and Meth. {\bf A453}, 606 (2000).
 \bibitem{jan02} S.~Janssen, dissertation University of Gie{\ss}en (2002). 
 \bibitem{geant3} R.~Brun, F.~Bruyant, A.C.~McPherson and P.~Zanarini, GEANT3
Users Guide, Data Handling Division DD/EE/84-1, CERN (1986).
% \bibitem{kru02} B.~Krusche et al., in preparation (2002).
 \end{references}
\end{document}